\RequirePackage{fix-cm}
\documentclass{svjour3}                     
\smartqed  
\usepackage{graphicx}
\usepackage{amsmath}
\usepackage[round]{natbib}
 \usepackage{subfig}
\usepackage{latexsym}
\newsavebox{\astrutbox}
\sbox{\astrutbox}{\rule[-5pt]{0pt}{20pt}}

\def\der#1#2{{\partial #1\over \partial #2}}

\def\be{\begin{equation}}
\def\ee{\end{equation}}
\def\bea{\begin{eqnarray}}
\def\eea{\end{eqnarray}}
\def\bse{\begin{subequations}}
\def\ese{\end{subequations}}
\def\bsea{\begin{subeqnarray}}
\def\esea{\end{subeqnarray}}
\def\({\left (}
\def\){\right )}
\def\[{\left [}
\def\]{\right ]}
\def\<{\left <}
\def\>{\right >}

\begin{document}

\title{On the physical mechanism of centrifugal-gravity wave resonant instability in swirling free surface rotating Polygons}

\titlerunning{}

\author{Ron Yellin-Bergovoy, Eyal Heifetz \& \\ Orkan M. Umurhan}


\institute{Ron Yellin-Bergovoy \at Department of Geosciences, Tel-Aviv University, Tel-Aviv, Israel\\
{\email yellinr@post.tau.ac.il}\ \and
Eyal Heifetz \at Department of Geosciences, Tel-Aviv University, Tel-Aviv, Israel\\
{\email eyalh@post.tau.ac.il} \and
Orkan M. Umurhan \at NASA Ames Research Center, Division of Space Sciences, Planetary Systems Branch, Moffett Field, CA, 94035 USA\\
{\email orkan.umurhan@gmail.com} 	}

\date{\today}

\maketitle




%
\begin{abstract}

We present an explicit analysis of wave-resonant instability of swirling flows inside fast rotating cylindrical containers. The linear dynamics are decomposed into the interaction between the horizontal inner centrifugal edge waves, the outer vertical gravity waves with the aim of understanding the dynamics of the centrifugal waves. We show how the far field velocity induced respectively by the centrifugal and the gravity waves affect each other's propagation rates and amplitude growth. We follow this with an analysis of the instability in terms of a four wave interaction, two centrifugal and two gravity ones, and explain why the resonant instability can be obtained only between a pair of two counter-propagating waves, one centrifugal and one gravity. 
Furthermore, a near resonant regime which does not yield instability is shown to result from a phase-locking configuration between a pair of a counter-propagating centrifugal wave and a pro-propagating gravity one, where the interaction affects the waves' propagation rates but not the amplitude growth.     

\end{abstract}
%
%
%

\section{Introduction}

\begin{figure}[]
\begin{center}  
                \fbox{\includegraphics[width=0.75\textwidth]{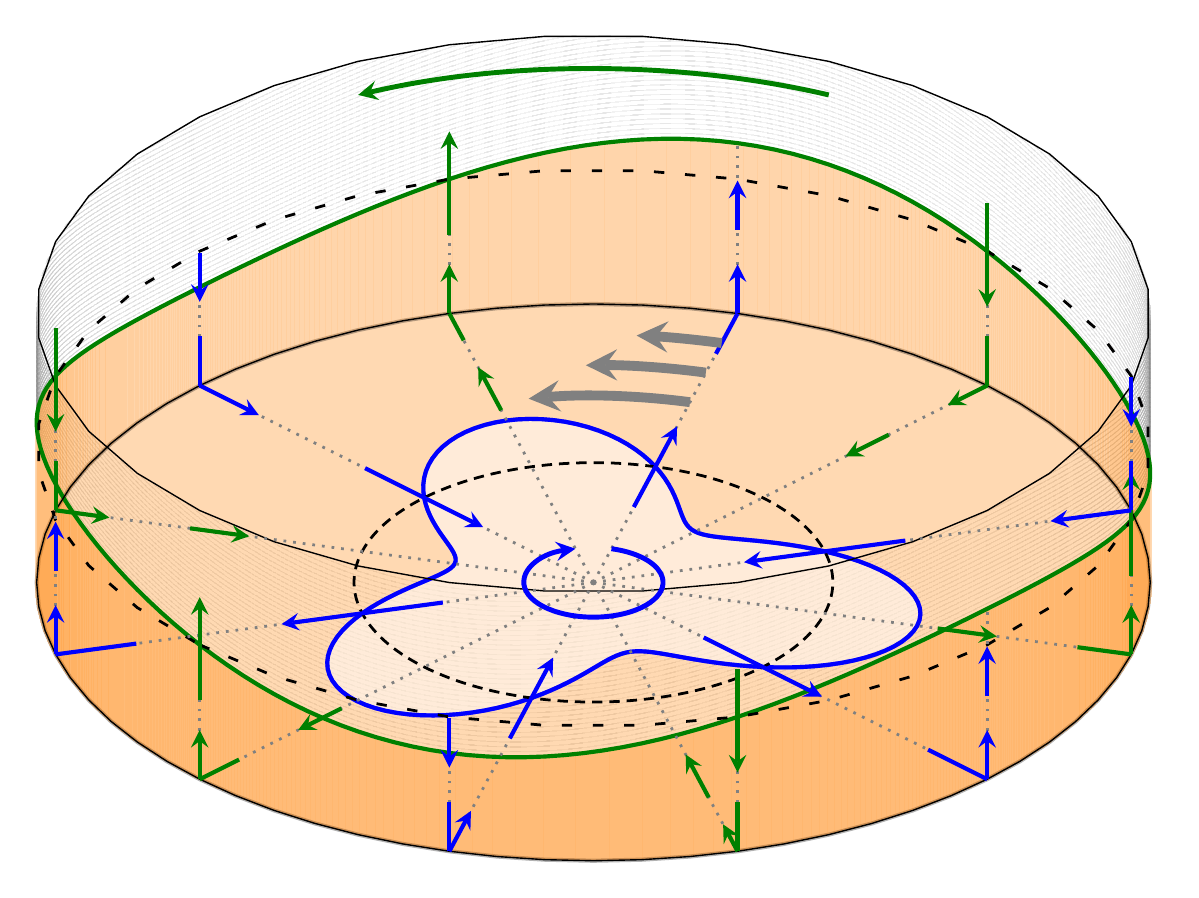}}
\caption{\small Resonant interaction between counter propagating gravity wave (green), and a counter propagating centrifugal wave (blue). The green and blue azimuthal arrows represent the waves direction of propagation and the radial/vertical arrows indicate the induced velocity field by each wave. The gray thick arrows illustrates the basic state flow velocity ($\bar V \propto{1\over r}$) and the black dashed lines show the interfaces of the unperturbed mean flow. In addition the white polygon in the middle of the cylinder represent the "dry patch" of the vortex.}
\label{fig:1}
\end{center}
\end{figure}

The emergence of steady polygonal patterns in swirling flows inside rotating cylindrical containers is both beautiful and intriguing. 
Recent work \citep{mougel2015waves,fabre2014generation,mougel2014waves,tophoj2013rotating} has
shown that at high rotation rates the mean flow rotates approximately as an irrotational vortex (denoted by Fabre and \citet{mougel2014waves}, hereafter FM14, as the ``Dry Potential'' regime). In this regime the observed polygonal patterns result mainly from resonant interaction between vertical-azimuthal gravity waves on the outer perimeter of the cylinder (at the top of the flow) and centrifugal horizontal-azimuthal waves on the cylinder surface, at the inner interface between the flow and the air (Fig.\ref{fig:1}). \citet{tophoj2013rotating}, hereafter TMBF, employed potential flow theory to simplify the dynamics and showed that their analysis captures the essence of the dynamics, a fact that has been confirmed by FM14. Nevertheless, the explicit propagation mechanism of the inner centrifugal waves as well as the interaction mechanism between the gravity and the centrifugal wave remains somewhat obscured. 

Tangent to that, there is a growing body of literature devoted to understanding various scenarios of shear instability in terms of interaction-at-a-distance between counter-propagating interfacial vorticity waves \citep[][to name a few]{hoskins1985use,baines1994mechanism,heifetz1999counter,harnik2008buoyancy,heifetz2015stratified}. In brief, the idea behind this way of thinking is that the phase relation between the wave's vorticity and displacement determines the direction of the wave propagation in isolation. In the presence of shear flow two interfacial waves with an oppositely signed vorticity-displacement relationship may remain phase-locked to each other if each wave propagates counter its local mean flow (viewed from the frame of reference of the averaged mean flow). By implementing vorticity inversion, that is by obtaining the velocity far field induced by the interfacial vorticity waves, one can formulate how each wave pushes the displacement of the other further. If the counter-propagating waves are phase locked this mutual amplification sustains and thus enables resonance instability.  For more details 
on this physical scenario the reader is referred to the review paper by \citet{carpenter2011instability}.     

Here, we wish to implement this wave action at-a-distance concept to the free surface swirling flow, while keeping the potential flow formulation of TMBF. As we will see, this analysis sheds light both on the nature of the inner horizontal centrifugal waves as well as on the resonant mechanism between the latter and the vertical gravity waves at the outer circumference of the cylinder. 

The rest of the paper is organized as follows. In Section 2 we formulate the problem setup and linearize the equations with respect to the mean flow. In Section 3 we investigate the propagation mechanism of the waves in isolation, where in Section 4 we write the explicit wave interaction equations and solve them for the resonance condition. We end by discussing our results and routes for future work.

\section{Formulation}

\subsection{Setup}

\begin{figure}[]
\begin{center}  
                \fbox{\includegraphics[width=0.75\textwidth]{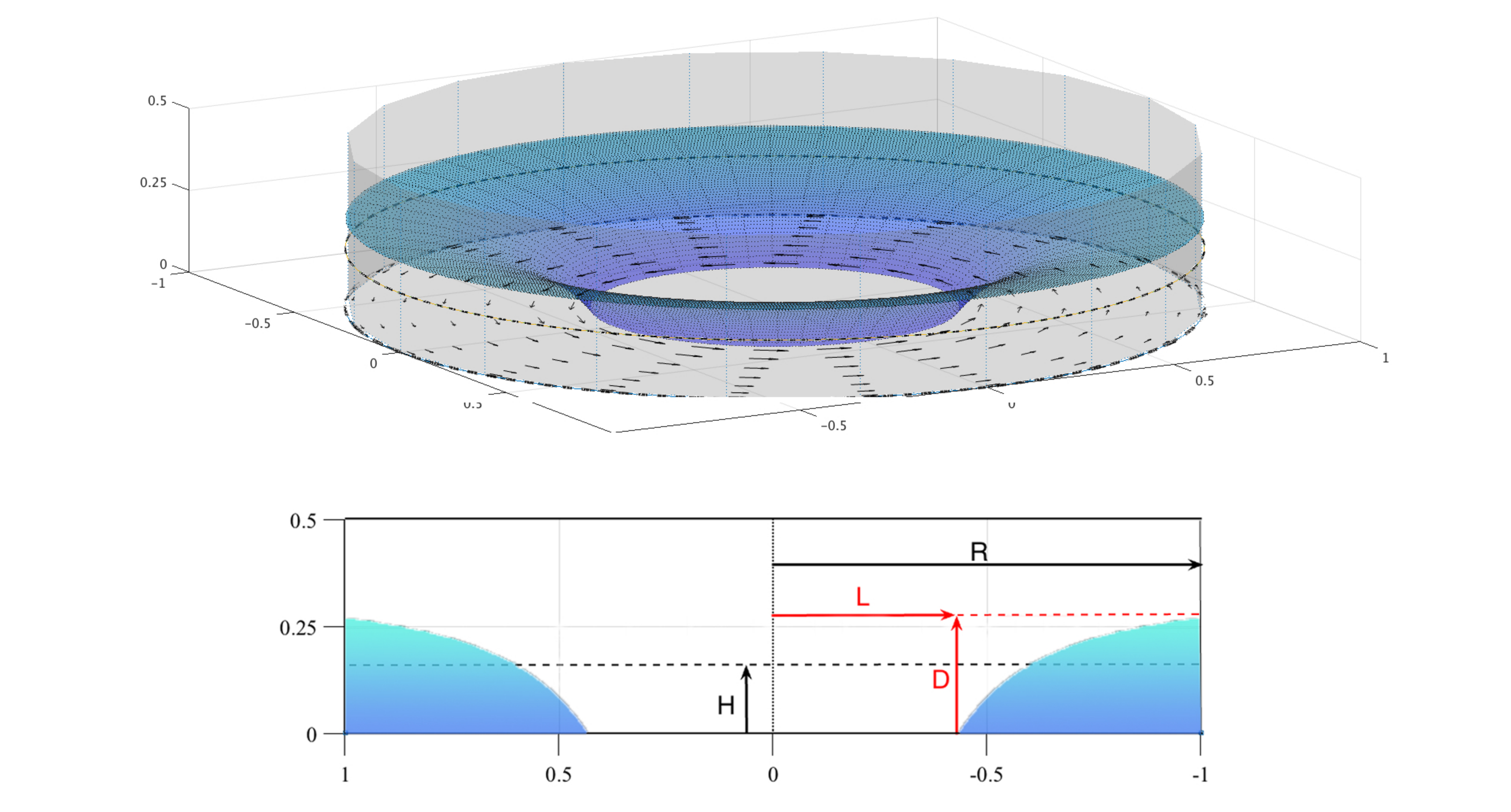}}
\caption{\small Side view (top) and radial-vertical cross-section (bottom) of the unperturbed equilibrium flow in a state of an irrotational vortex, scaled by the cylinder radius $R$. 
$H$ is the fluid height at rest, $D$ is the maximum height of the rotated flow and $L$ is its inner radius.}
\label{fig:2}
\end{center}
\end{figure}

We consider a potential flow in a rotating cylindrical tank. The governing momentum equations in the radial ($r$); azimuthal ($\theta$); and height ($z$) coordinates for the respective ${\bf v} = (u,v,w)$ velocity components, can be written in the inertial frame as: 
\be
\label{m1_inert}
\frac{Du}{Dt} = \frac{v^2}{r}  -\frac{1}{\rho}\frac{\partial p}{\partial r},
\ee
\be
\label{m2_inert}
\frac{Dv}{Dt} = -\frac{uv}{r} -\frac{1}{\rho r}\frac{\partial p}{\partial \theta},
\ee
\be
\label{m3_inert}
\frac{Dw}{Dt} = -g  -\frac{1}{\rho}\frac{\partial p}{\partial z}.
\ee
Here the constant density is $\rho$, $p$ is the pressure, $g$ is gravity and $t$ denotes time.
The material derivative is $\frac{D}{Dt} \equiv \der{}{t} + u\der{}{r} +{v \over r}\der{}{\theta} + w\der{}{z}$. 
Following TMBF we assume that the azimuthally independent unperturbed equilibrium flow (denoted by overbars) is an irrotational vortex with circulation 
$\Gamma$, sketched in Fig. \ref{fig:2}:
\be 
\label{eq4}
{\overline U} = 0\,; \hspace{0.25cm} {\overline V} = {\overline \Omega} r = {\Gamma \over 2\pi r}\,; \hspace{0.25cm} {\overline W} = 0\, .
\ee
Substituting the mean flow solution into the momentum equations (\ref{m1_inert}-\ref{m3_inert}) yields the partial differential equations governing the unperturbed pressure gradient force,
\be
\label{eq5}
\frac{1}{\rho}\der{}{r}{\overline P}(r,z) = {{\overline V}^2 \over r} = {\overline \Omega}^2 r \,; \hspace{0.25cm}
\frac{1}{\rho}\der{}{z} {\overline P}(r,z)= -g\, .
\ee
Since the pressure is constant on the free surface of the flow, the surface height $Z_s$ satisfies:
\be
\label{eq6}
{Z_s(r) \over H} = {F^2 \over 2}\[\(R\over L\)^2 -\(R\over r \)^2\]\, , 
\ee
where $L$ and $R$ are respectively the inner and outer radii of the irrotational vortex and the subscript $s$ denotes the free surface. $H$ is the fluid height at rest and 
\be
\label{eq7}
F^2 = {{\overline V}^2(R) \over gH},
\ee
is the square of the Froude number for this setup. We note that $F = F_p/\sqrt{A}$, where $F_p \equiv \Omega(R) \sqrt{R/g}$ is the ``plate'' Froude number, as defined by FM14,
and $A = H/R$ is the aspect ratio of the fluid at rest. 
Mass conservation implies as well that  $D = Z_s(R)$ satisfies:
\be
\label{eq8}
{D \over H} = 1 +F^2\ln\({R\over L} \). 
\ee
Equating \eqref{eq6} with  \eqref{eq8} for $r=R$, with the aid of  \eqref{eq4}, gives expressions for $F$, $\Gamma$ and $D$ in terms of $(H,R,L)$ 
(see TMBP for more details, please note however the different notation).
FM14 showed that for $F_p > 2.5$ and $A =0.3$, that is for $F>5$, the potential flow dynamics introduced by TMBF provides a very good approximation to the ``Dry Potential'' dynamics. For comparison with the results of TMBF, we adopt an aspect ratio of $A =0.276$.

\subsection{Linearized dynamics}

Linearization of the momentum equations (\ref{m1_inert}-\ref{m3_inert}) with respect to the basic state of  \eqref{eq4} and  \eqref{eq5} yields,
\be
\label{eq9}
\frac{D_L u'}{Dt} = 2{\overline \Omega}v'  -\frac{1}{\rho}\frac{\partial p'}{\partial r},
\ee
\be
\label{eq10}
\frac{D_L v'}{Dt} =  -\frac{1}{\rho r}\frac{\partial p'}{\partial \theta},
\ee
\be
\label{eq11}
\frac{D_L w'}{Dt} = -\frac{1}{\rho}\frac{\partial p'}{\partial z}.
\ee
where $\frac{D_L }{Dt} = \der{}{t}+ {\overline \Omega}\der{}{\theta}$ is the linearized material derivative. 

The free surface pressure $p_s$ is assumed to be materially conserved by the fluid parcels occupying the surface, thus ${Dp_s \over Dt} = 0$. Decomposing $p_s$ to its balanced and perturbed components: $p_s(r,\theta,z,t) = {\overline P}_s(r,z) + p'_s(r,\theta,z,t)$, this material conservation, together with \eqref{eq5}, gives,
\be
\label{eq12}
{D p'_s \over Dt} = - {D {\overline P}_s \over Dt} = \rho\[- {\overline \Omega}^2 r u' +  g  w'\]_s \, ,
\ee
hence under linearization
\be
\label{eq13}
p'_s = \rho\[- {\overline \Omega}^2 r \chi' +  g \eta' \]_s \, ,
\ee
in which $(\chi', \eta')$ are the radial and vertical displacements of the perturbed free surface,
formally written in linearized form as
\be
\frac{D_L \chi'}{D t} = \Big[u'\Big]_{s}, \qquad
\frac{D_L \eta'}{D t} = \Big[w'\Big]_{s}.
\ee
Specifically around $Z_s(L) = 0$
and $Z_s(R) = D$ we have the following, 
\be
\label{eq14}
\[{p_s}= - \rho {\overline \Omega}_{L}^2 L \chi\]_{(L,0)} \,; \hspace{0.25cm}
\[{p_s} =  \rho g\eta\]_{(R,D)}\, ,
\ee
where hereafter we omit the primes for perturbations.
Substituting these expressions back in (\ref{eq10}) reveals,
\be
\label{eq15}
\[\frac{D_L v}{Dt} = {\overline \Omega}_{L}^2 \frac{\partial \chi}{\partial \theta}\,; \hspace{0.25cm}
\frac{D_L \chi}{Dt} = u \hspace{0.1cm} \Rightarrow \hspace{0.1cm} 
\frac{D^2_L v}{Dt^2} = {\overline \Omega}_{L}^2 \frac{\partial u}{\partial \theta}\]_{(L,0)}\, ,
\ee
\be
\label{eq16}
\[\frac{D_L v}{Dt} = -{g \over R}\frac{\partial \eta}{\partial \theta}\,; \hspace{0.25cm}
\frac{D_L \eta}{Dt} = w  \hspace{0.1cm} \Rightarrow \hspace{0.1cm} 
\frac{D^2_L v}{Dt^2} = -{g \over R}\frac{\partial w}{\partial \theta}\]_{(R,D)}\,.
\ee

\section{Interfacial wave dynamics}
\subsection{Vortex sheet representation in potential flow}

In the reported experiments, the swirling flow inside the container is surrounded by the ambient air. Hence, the vorticity perturbation at the interfaces can be estimated as,
\be
\label{eq17}
\[(\nabla\times{\bf u})\cdot\hat {\bf z} = \(\der{(rv)}{r} -  {1\over L}\der{u}{\theta}\)\]_{(L,0)} = \(v_f - v_a\)\delta(r-L) \, ,
\ee
\be
\label{eq18}
\[(\nabla\times{\bf u})\cdot\hat {\bf r} = \(-\der{v}{z} +  {1\over R}\der{w}{\theta}\)\]_{(R,D)} = \(v_f - v_a\)\delta(z-D) \, ,
\ee
where the subscripts $(f,a)$ represent respectively the swirling flow and the ambient air.
Physically, this vorticity $\delta$-function is generated by the baroclinic torque since both the density and the pressure change abruptly across the interfaces between the flow and the ambient air \citep[see a detailed analysis of the mechanism in][] {heifetz2015stratified}. The velocity far field induced by each interface can be obtained by vorticity inversion, i.e., by finding the Green function associated with those interfacial vorticity $\delta$-functions. 

Under the potential flow approximation of TMBF the ambient air dynamics ($v_a$)is neglected and therefore the tangential velocity perturbation ($v_f$) at the interface represents the vorticity $\delta$-function there. Hence, the induced velocity field by each interface can be obtained by finding the velocity potential inducing zero tangential velocity signature on the opposed interface. This is done explicitly in the next subsection.


\subsection{Velocity splitting}

We consider the perturbation velocity potential in the form of $\phi = \Phi(r,z,t)e^{im\theta}$, so that ${\bf v} = \nabla\phi$, and $\nabla^2 \phi = 0$.
Following TMBF we assume that both the normal and the tangential components of the velocity are continuous (but not necessarily zero) at $(r,z) = (R,0)$ :
\be
\label{eq19}
\[u = \der{\phi}{r}\]_{r \rightarrow R}^{z=0} =  \[w = \der{\phi}{z}\]_{r=R}^{z \rightarrow 0}\, ; \hspace{0.5cm}
v = \[{1\over r}\der{\phi}{\theta}\]_{r \rightarrow R}^{z=0} = \[{1\over r}\der{\phi}{\theta}\]_{r=R}^{z \rightarrow 0}\, .
\ee
Next we decompose the potential to the parts attributed to the bottom $(L, 0)$ and the top $(R,D)$ interfacial waves: $\phi =  \phi_B + \phi_T$, so that
the bottom (top) wave induces zero tangential velocity at the bottom (top) one. This implies: 
\be
\label{eq20}
{\phi_B}_{(L, 0)} = -i{L\over m}{v_B}_{(L, 0)}\, ; \hspace{0.25cm} {\phi_B}_{(R,D)} = 0\, , \hspace{0.25cm}
{\phi_T}_{(L, 0)} = -i{R\over m}{v_T}_{(R, D)}\, ; \hspace{0.25cm} {\phi_T}_{(L,0)} = 0\, .
\ee 
The solution of the Laplace equation together with the boundary conditions of (\ref{eq19},\ref{eq20}) yields:
\be
\label{eq21}
{\phi_B} = {-i{L\over m}{v_B}_{(L, 0)} \over \[ e^{-2mD\over R}\({L\over R}\)^m - \({L\over R}\)^{-m}\]}
\begin{cases}
\[e^{-2mD\over R}\({r\over R}\)^m -  \({r\over R}\)^{-m}  \]            ,& (L \le r \le R, \, z=0)\\
\[e^{-2mD\over R} e^{mz\over R} - e^{-mz\over R}\]     ,& (r=R, \,  0 \leq z \leq D)
\end{cases}\, ,
\ee
\be
\label{eq22}
{\phi_T} = {-i{R\over m}{v_T}_{(R, D)} \over \[ e^{mD\over R} - e^{-mD\over R}\({L\over R}\)^{2m}\]}
\begin{cases}
\[\({r\over R}\)^m -  \({L\over R}\)^{2m}\({r\over R}\)^{-m}  \]            ,& (L \le r \le R, \, z=0)\\
\[e^{mz\over R} -  e^{-mz\over R}\({L\over R}\)^{2m}\]     ,& (r=R, \,  0 \leq z \leq D)
\end{cases}\, .
\ee
Defining
\be
\label{eq23}
\alpha \equiv e^{-mD\over R}\({L\over R}\)^{m}\, ; \hspace{0.25cm} 
\beta \equiv {(1 + \alpha^2) \over (1 - \alpha^2)}\, ; \hspace{0.25cm} 
\gamma \equiv  {\alpha \over (1 - \alpha^2)}\, ,
\ee
then at the interfaces \eqref{eq21} and \eqref{eq22} give:
\be
\label{eq24}
{u_B}_{(L,0)} = i\beta{v_B}_{(L, 0)}\, ; \hspace{0.25cm}
{u_T}_{(L, 0)} = -2i{R\over L}\gamma{v_T}_{(R, D)}\, ,
\ee
\be
\label{eq25}
{w_B}_{(R, D)} = 2i{L\over R}\gamma{v_B}_{(L, 0)}\, ; \hspace{0.25cm}
{w_T}_{(R, D)} = -i\beta{v_T}_{(R, D)}\, .
\ee

\begin{figure}[]
\begin{center}  
                \fbox{\includegraphics[width=0.75\textwidth]{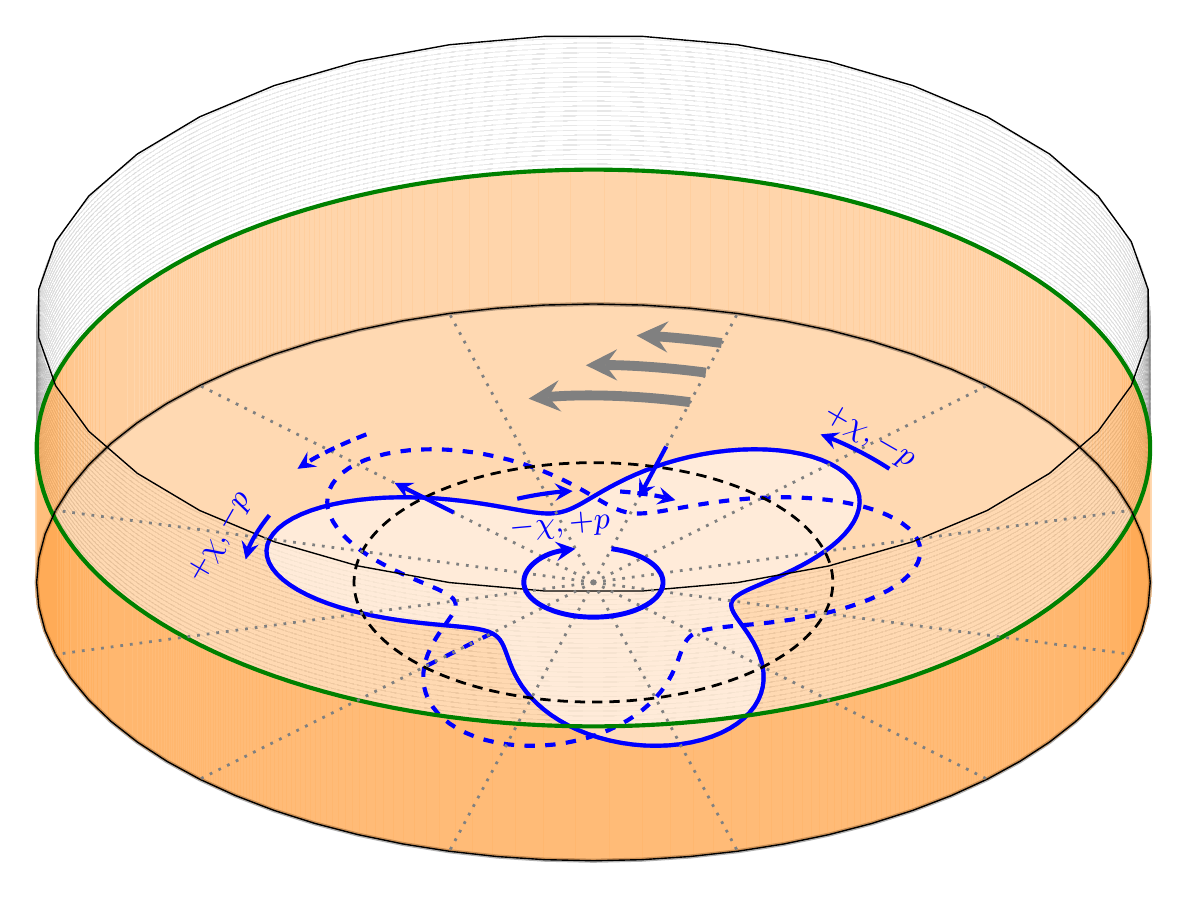}}
\caption{\small Centrifugal wave counter propagation mechanism  on the cylinder horizontal surface, at the inner interface between the flow and the air. Solid lines and arrows represent the fluid displacement and velocity field at current time $t$, where the dashed ones represent these fields at time $(t+T/4)$ (where $T$ is the wave period). Radial displacement is indicated by ($\chi$) and the pressure anomaly by ($\pm p$).}
\label{fig:3}
\end{center}
\end{figure} 

\subsection{Interfacial wave propagation}

The complete interfacial dynamics can be obtained by substituting \eqref{eq24} and \eqref{eq25} in \eqref{eq15} and \eqref{eq16}:
\be
\label{eq26}
\[\frac{D^2_L v_B}{Dt^2} = {\overline \Omega}_{L}^2 \frac{\partial }{\partial \theta}\(u_B +u_T\)\]_{(L, 0)}\, ; \hspace{0.25cm}
\[\frac{D^2_L v_T}{Dt^2} = -{g\over R} \frac{\partial }{\partial \theta}\(w_B +w_T\)\]_{(R, D)} \, .
\ee
Nonetheless, the velocity splitting provides us a natural way to consider the dynamics of each interface in isolation by looking only at the self interacting terms: 
\be
\label{eq27}
\[\frac{D^2_L v_B}{Dt^2} = {\overline \Omega}_{L}^2 \frac{\partial u_B}{\partial \theta}\]_{(L, 0)} \, ; \hspace{0.25cm}
\[\frac{D^2_L v_T}{Dt^2} = -{g\over R} \frac{\partial w_T}{\partial \theta}\]_{(R, D)} \, .
\ee
Assuming a wave-like solution of the form of $e^{i(m\theta -\omega t)}$, \eqref{eq24}, \eqref{eq25} and \eqref{eq27} give the dispersion relations for the centrifugal and gravity waves:
\be
\label{eq28}
\omega_c^{\pm} = m{\overline \Omega}_{L}\(1\pm \sqrt{\beta \over m}\) \, ; \hspace{0.25cm}
\omega_g^{\pm} = m{\overline \Omega}_{R}\[1\pm \(\sqrt{R\beta \over HFm}\)\] \, ,
\ee
which are the same dispersion relations (with different notations) obtained by TMBF. The subscripts $(c,g)$ denote the centrifugal and gravity waves, where the 
$(+,-)$ superscripts refer to the waves whose phase propagation is (larger, smaller) than the mean velocity at the interface.  

\begin{figure}[]
\begin{center}  
                \fbox{\includegraphics[width=0.75\textwidth]{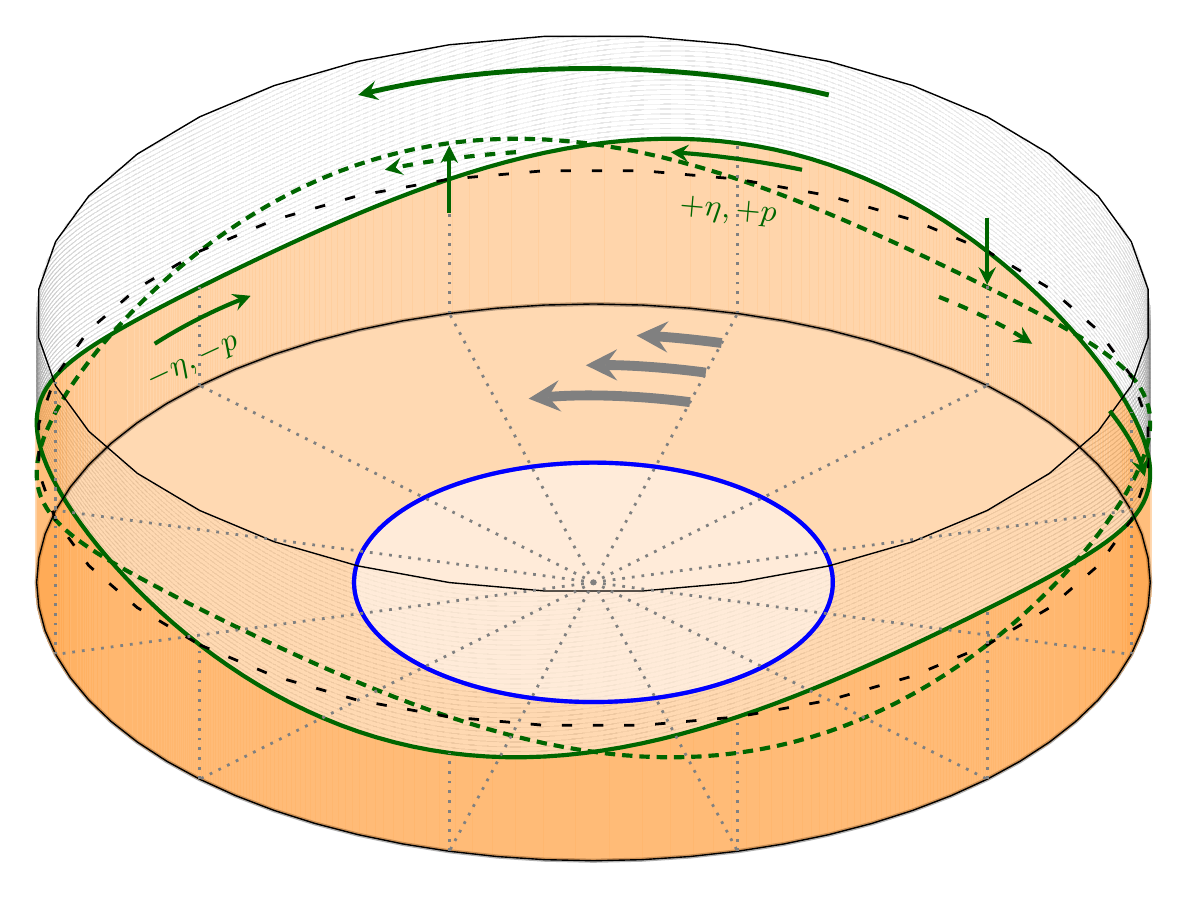}}
\caption{\small Gravity wave propagation mechanism. Notations and symbols are like in Fig .\ref{fig:3}, except that ($\pm\eta$) represents vertical wave displacement. We denote this wave as counter-propagating since it propagates counter the mean flow, measured from a frame of reference moving with the mean azimuthal mean flow.}
\label{fig:4}
\end{center}
\end{figure}

We aim to understand the propagation mechanism of these waves, especially the centrifugal ones. Toward this end we look at their structure at the interfaces, using (\ref{eq14}-\ref{eq16}):
\be
\label{eq29}
\[\(\chi,\, u,\, p\)_c^{\pm} = 
\(\mp {1\over {\overline \Omega}_{L}}\sqrt{\beta \over m},\, i\beta\, , \pm\rho L {\overline \Omega}_{L}\sqrt{\beta \over m}\, \)v_c^{\pm}\]_{(L, 0)}\, , 
\ee  
\be
\label{eq30}
\[\(\eta,\, w,\, p\)_g^{\pm} = 
\(\pm \sqrt{R\beta \over g m},\, -i\beta\, , \pm\rho \sqrt{Rg\beta \over  m}\, \)v_g^{\pm}\]_{(R, D)}\, .
\ee 
In Fig. \ref{fig:3} we sketch the propagation mechanism of the {counter-propagating centrifugal wave ($c^-$)} . As indicated from (\ref{eq29}) the radial displacement $\chi$ and the tangential velocity $v$ are in phase. Since outward radial displacement (positive $\chi$) retreats the flow from the mean interface it decreases the pressure anomaly there, hence 
($\chi \propto -p \propto v$). As discussed previously, $v$ represents the vorticity $\delta$-function at the interface (which is positive for 
counter-clockwise circulation) and indeed $u$ is lagging $v$ by a quarter of a wavelength to generate together counter-clockwise rotation, 
in phase with $\chi$.
At the wave nodes the radial displacement, the tangential velocity and the pressure anomalies are all zero. 
The clockwise propagation mechanism of the wave (with respect to the mean flow there which is counter-clockwise) becomes intuitive now: the non zero radial velocity $u$ at the nodes translates the radial displacement anomalies (and hence the pressure), whereas the tangential component of the pressure gradient force at the nodes accelerates the flow, thus translating $v$ in concert. By flipping the sign relations to  ($-\chi \propto p \propto v$) it is straightforward to illustrate the propagation mechanism of the pro-propagating centrifugal wave 
($c^+$). Although the gravity wave propagation mechanism is well known it is interesting to point out that it can be explained as well in a similar fashion {\citep[for more details see][]{harnik2008buoyancy}}. Fig. \ref{fig:4} demonstrates the propagation mechanism of the { counter propagating gravity wave ($g^+$)}  where ($\eta \propto p \propto v$) at the upper interface.

\section{Wave resonance}

\subsection{Wave interaction equations}

Equations (\ref{eq24}-\ref{eq26}) contain all the required information to solve the instability problem. However since each interface supports two waves, the solution by itself does not provide explicit understanding on how the four waves interact. Here we follow the method suggested by \citet{harnik2008buoyancy} to obtain explicit equations for the interfacial wave dynamics, expressed solely in terms of the waves' displacements across the interfaces.

\begin{figure}[]
\begin{center}  
                \fbox{\includegraphics[width=0.75\textwidth]{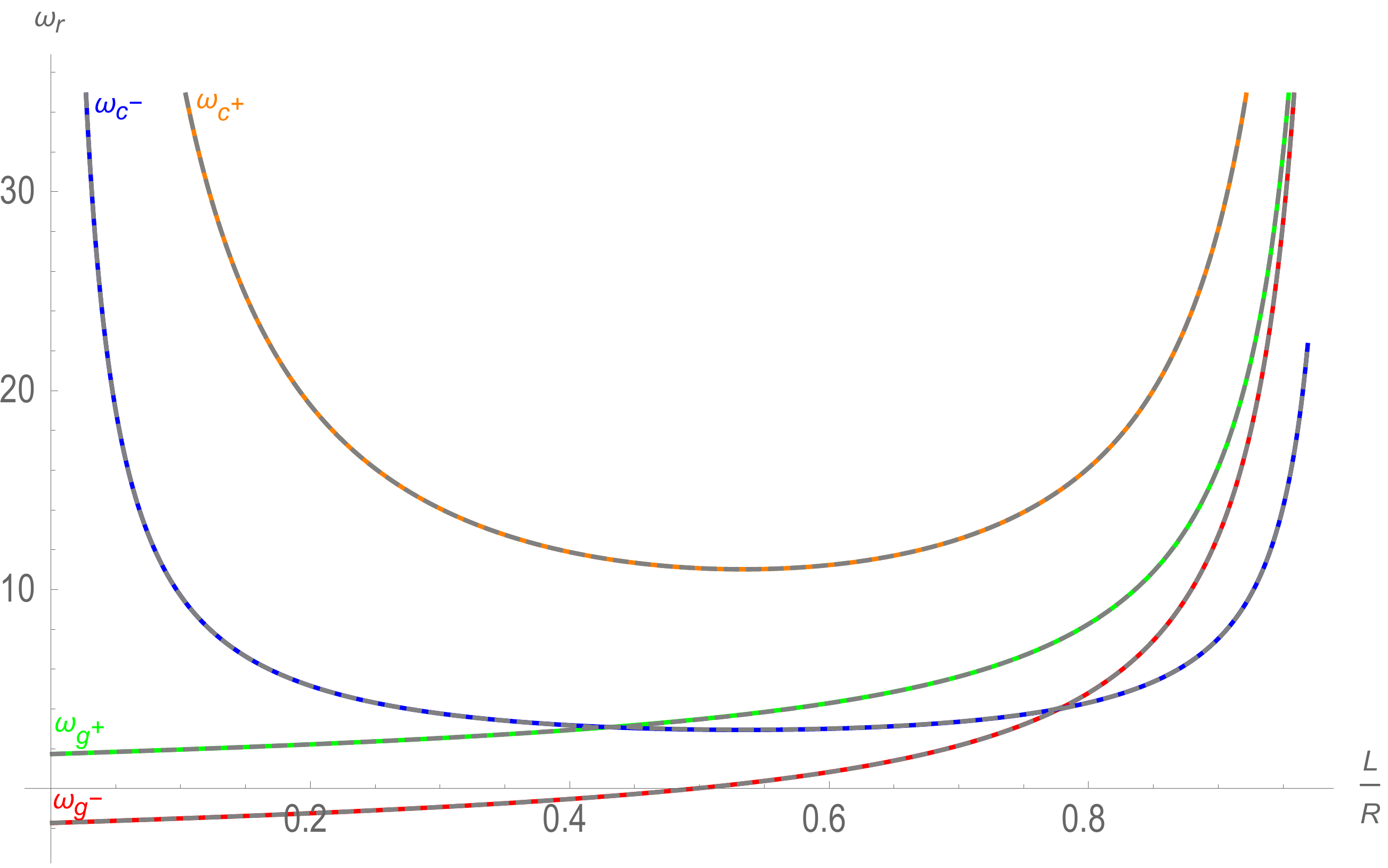}}
\caption{\small The real part of dispersion relation frequencies as a function of ${L}/{R}$ for ${H}/{R}=0.276$ and $m=3$. The solid gray lines are of the four branches of 
$(\omega_{NM_r})$, whereas the four solid colored dashed lines represent the pro-propagating (orange) and  counter-propagating (blue) centrifugal waves and the counter-propagating (green) and pro-propagating (red) gravity waves frequencies. All frequencies are scaled by $\sqrt{\frac{g}{R}}$ as in TMBF.}
\label{fig:5}
\end{center}
\end{figure}

Toward this end we express the perturbations as a sum of individually propagating interfacial waves,
\be 
\label{eq31}
\phi_B = \phi_c^+ + \phi_c^- \, ; \hspace{0.1cm}
\phi_T = \phi_g^+ + \phi_g^- \, ; \hspace{0.1cm}  
{\chi_{}}_B = \chi_c^+ + \chi_c^- \, ; \hspace{0.1cm}
{\eta_{}}_T =  \eta_g^+ + \eta_g^- \, ,  
\ee
where, as indicated by \eqref{eq29} and \eqref{eq30}, we have,
\be
\label{eq32}
\[\chi_c^{\pm} = \mp \({1\over {\overline \Omega}_{L}}\sqrt{\beta \over m}\)v_c^{\pm}\]_{(L, 0)}\, \, ; \hspace{0.25cm}
\[\eta_g^{\pm} = \pm \(\sqrt{R\beta \over g m}\)v_g^{\pm}\]_{(R, D)}\, .
\ee 
Substituting \eqref{eq31} and \eqref{eq32} into \eqref{eq15} and \eqref{eq16} yields,
\be
\label{eq33}
\[\(\der{}{t} + i\omega_c^{\pm}\)\chi_c^{\pm} = {1\over 2}u_T\]_{(L, 0)}\, \, ; \hspace{0.25cm}
\[\(\der{}{t} + i\omega_g^{\pm}\)\eta_g^{\pm} = {1\over 2}w_B\]_{(R, D)}\, \, .
\ee
The interpretation of \eqref{eq33} is straightforward. Without interaction (when the RHS is zero) the intrinsic wave frequencies of \eqref{eq28} are recovered.
With interaction, the induced far field velocity by the perturbation of a given interface is equi-partitioned between the two interfacial waves at the opposed interface. Writing explicitly $\[u_T\]_{(L, 0)}$ in terms of $\( \eta_g^+, \eta_g^-\)$ and $\[w_B\]_{(R, D)}$ in terms of $\(\chi_c^+, \chi_c^-\)$ we obtain,
\be
\label{eq34}
\(\der{}{t} + i\omega_c^{\pm}\)\chi_c^{\pm} = -i\sigma_T\(\eta_g^+ - \eta_g^-\)  \,  ; \hspace{0.25cm}
\(\der{}{t} + i\omega_g^{\pm}\)\eta_g^{\pm} = -i\sigma_B\(\chi_c^+ - \chi_c^-\)  \, ,
\ee
where the interaction coefficients are,
\be
\label{eq35}
\sigma_T \equiv \gamma \sqrt{mg R \over \beta L^2}  \,  ; \hspace{0.25cm}
\sigma_B \equiv \gamma {\overline \Omega}_{L}{L\over R}\sqrt{m \over \beta } \, .
\ee
The four equations in \eqref{eq34} describe the explicit interaction between the four interfacial waves in terms of their local displacement at the interfaces. Substituting normal mode solutions of the form $e^{i(m\theta-\omega_{NM} t)}$ into (35) we obtain the modal dispersion relation. The real part $(\omega_{NM_r})$ solutions are shown in figure \ref{fig:5}  as a function of $(L/R)$ at azimuthal wavenumber $m=3$. 
The curves are identical to the ones presented in figure 2a of TMBF. The waves' frequencies without interaction $(\omega_c^{\pm},\omega_g^{\pm})$, are presented as well in this figure and nearly coincide with the corresponding normal mode frequencies. This indicates that in most regions each normal mode is dominated by the propagation of a single interfacial wave. 

\subsection{Counter-propagating wave dynamics}

\begin{figure}%
    \centering
    \subfloat[]{{\fbox{\includegraphics[width=10cm]{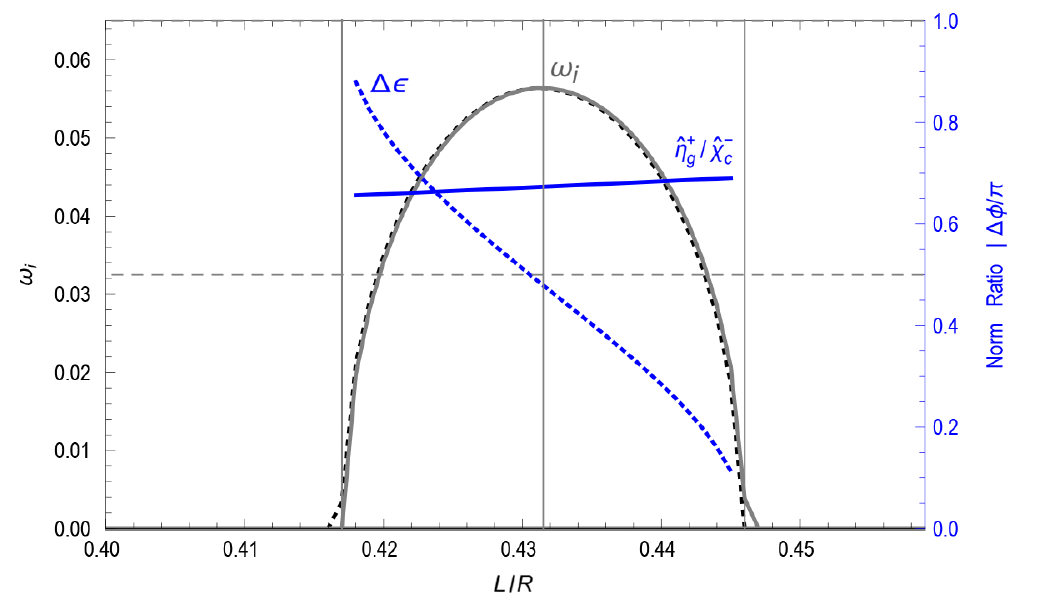}}
    \label{fig:6a} }}
    \qquad
    \subfloat[]{{\fbox{\includegraphics[width=10cm]{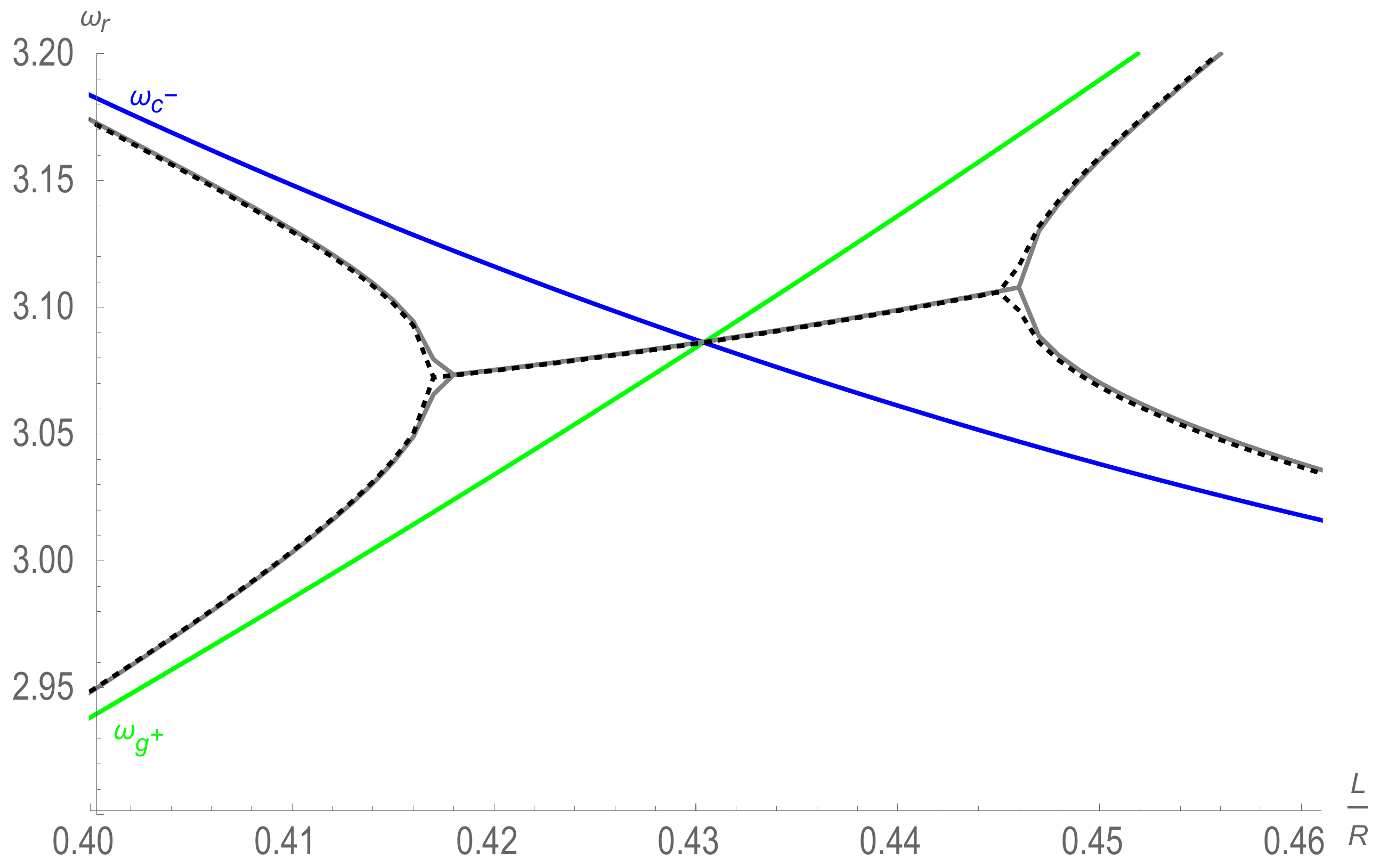} }
    \label{fig:6b}}}%
    \caption{\small Comparison between the unstable normal mode solution and the two counter-propagating waves approximation for ${H}/{R}=0.276$ and m=3. (a) The solid gray and the black dashed lines represent respectively the growth rate ${\omega_{NM}}_i$ obtained from the full solution and from the two-wave system. The phase difference between the counter-propagating waves $(\Delta \epsilon = (\epsilon_g^{+} - \epsilon_c^{-})$ and their amplitude  ratio $({{\hat \eta}_g^{+}/{\hat \chi}_c^{-}})$ in the unstable regime are indicated by the dashed and  solid blue lines, respectively. (b) Blow up of Figure. 5 in the unstable regime. Solid gray and black dashed lines represent respectively ${\omega_{NM}}_r$, obtained from the full solution and from the two-wave system. The frequencies $\omega_c^{-}$ and $\omega_g^{+}$ of the counter propagation waves in isolation are indicated as well by the blue and the green solid lines.}
    \label{fig:6}
\end{figure}

The positive branch of the imaginary frequency ($\omega_{NM_i} > 0$), indicating instability, is calculated from \eqref{eq34} and shown in figure \ref{fig:6a}. It is identical to the one in figure 2c of TMBF. A blow-up of figure \ref{fig:5} focusing oupon this region of instability is presented in figure \ref{fig:6b} and we confirm that it is also identical to figure 2b of TMBF.

We expect that the instability results from the resonance between the waves whose intrinsic frequencies (without interaction) are close to one another. 
Since ${\overline \Omega}_{L} > {\overline \Omega}_{R}$, \eqref{eq28} suggests that those waves are of $\(\chi_c^-,  \eta_g^+\)$. In the framework of the average mean frequency 
${\overline \Omega}_{M} \equiv {1\over 2}\({\overline \Omega}_{L} +{\overline \Omega}_{R}\)$, they propagate counter their local mean flow. We therefore approximate the solution by excluding the pro-propagating waves $\(\chi_c^+,  \eta_g^-\)$. As a result \eqref{eq34} is simplified to a two wave interaction dynamics: 
\be
\label{eq36}
\(\der{}{t} + i\omega_c^{-}\)\chi_c^{-} \approx -i\sigma_T\eta_g^+ \,  ; \hspace{0.25cm}
\(\der{}{t} + i\omega_g^{+}\)\eta_g^{+} \approx i\sigma_B\chi_c^-  \, .
\ee
The two solutions of \eqref{eq36} are then calculated and displayed in figure \ref{fig:6}(a,b). They are in a very good agreement with the full 4-wave modal solution, providing nearly exact values for the normal mode growth rates. Indeed, within the instability region, the amplitude ratios between the pro-propagating and the counter-propagating waves $(|\chi_c^{+}/\chi_c^{-}|, |\eta_g^{-}/\eta_g^{+}|)$ are calculated and found to be less than $2\%$ (not shown here). 

Being the essential players in the instability mechanism, we wish therefore to examine more closely the nature of interaction between the counter-propagating waves.  
We can write the wave displacements in terms of their amplitude and phases,  
\be 
\label{eq37}
\chi_c^{-} = {\hat \chi}_c^{-}(t)e^{i\[m\theta + \epsilon_c^{-}(t)\]}\,  ; \hspace{0.25cm}
\eta_g^{+} = {\hat \eta}_g^{+}(t)e^{i\[m\theta + \epsilon_g^{+}(t)\]}\, 
\ee
and substitute back in \eqref{eq36} to obtain equations for their instantaneous growth rates and frequencies,
\be
\label{eq38}
{{\dot  {\hat \chi}}_c^{-}\over {\hat \chi}_c^{-}} = {{\hat \eta}_g^{+}\over{\hat \chi}_c^{-}}\sigma_T\sin{\Delta \epsilon}\,  ; \hspace{0.25cm}
{{\dot  {\hat \eta}}_g^{+}\over  {\hat \eta}_g^{+}} = {{\hat \chi}_c^{-}\over {\hat \eta}_g^{+}}\sigma_B\sin{\Delta \epsilon}\, ,
\ee
\be
\label{eq39}
-{\dot \epsilon}_c^{-} = \omega_c^{-} + {{\hat \eta}_g^{+}\over{\hat \chi}_c^{-}}\sigma_T\cos{\Delta \epsilon}\,  ; \hspace{0.25cm}
-{\dot \epsilon}_g^{+} = \omega_g^{+} - {{\hat \chi}_c^{-}\over {\hat \eta}_g^{+}}\sigma_B\cos{\Delta \epsilon}\, ,
\ee
where $\Delta \epsilon \equiv (\epsilon_g^{+} - \epsilon_c^{-})$ is the displacement phase difference between the waves.
In figure 1 the waves are sketched with a phase difference of $\Delta \epsilon = \pi/2$. As indicated by eq. \eqref{eq38} this is the optimal configuration for instantaneous growth as the normal velocity, induced by each wave on the other, acts to increase the other wave's amplitude. Equation set \eqref{eq39} also indicates that in this configuration the wave interaction does not affect the intrinsic phase speeds of the waves. We can think of two other extreme cases, when $\Delta \epsilon = 0$ and $\Delta \epsilon = \pi$. In both cases eq. \eqref{eq38} shows that the wave interaction does not lead to amplitude growth. When $\Delta \epsilon = 0$ the wave displacements are in phase but the normal velocity induced by each other are anti-phased. As a result, the waves' propagation speeds are reduced, hence the waves hinder each other's propagation with respect to the mean flow. Consequently, the instantaneous frequency 
$-{\dot \epsilon}_c^{-}$, becomes more positive whereas $-{\dot \epsilon}_g^{+}$, becomes more negative as a result of the interaction. When $\Delta \epsilon = \pi$ the waves are 
anti-phased in terms of their displacement but in phase in terms of their normal velocity. This helps the wave to propagate counter the mean flow. For any phase relation in the 
range of $(0< \Delta \epsilon < \pi/2)$ the wave interaction makes the amplitudes to grow and hinder the counter-propagation rate. In the range of $(\pi/2< \Delta \epsilon < \pi)$
the waves amplify each other and help each other to counter-propagate against the average mean flow.

\subsection{Normal modes instability in terms of counter-propagating wave resonance}

Normal mode are the solutions of eqs. \eqref{eq38} and \eqref{eq39}, where, by definition, both waves experience the same growth rate (${\omega_{NM}}_i$) and are phase-locked to propagate with the same frequency (${\omega_{NM}}_r$): 
\be
\label{eq40}
{\omega_{NM}}_i = {{\dot  {\hat \chi}}_c^{-}\over {\hat \chi}_c^{-}} = {{\dot  {\hat \eta}}_g^{+}\over  {\hat \eta}_g^{+}}\,  ; \hspace{0.25cm}
{\omega_{NM}}_r = -{\dot \epsilon}_c^{-} = -{\dot \epsilon}_g^{+}\, .
\ee
These requirements determine the waves' amplitude ratio and phase difference: 
\be
\label{eq41}
{{\hat \eta}_g^{+}\over{\hat \chi}_c^{-}} = \sqrt{\sigma_B\over \sigma_T}\,  ; \hspace{0.25cm}
{\cos\Delta \epsilon} = {\omega_g^{+} - \omega_c^{-} \over 2 \sqrt{\sigma_B \sigma_T}}
\ee
as well as the explicit expressions for the normal mode growth rate and propagation frequency (figure \ref{fig:6}):
\be
\label{eq42}
{\omega_{NM}}_i = {1\over 2}\sqrt{\[4\sigma_B \sigma_T - (\omega_g^{+} - \omega_c^{-})^2\]}\,  ; \hspace{0.25cm}
{\omega_{NM}}_r = {1\over 2}(\omega_g^{+} + \omega_c^{-})\, .
\ee
The phase difference $(\Delta \epsilon = (\epsilon_g^{+} - \epsilon_c^{-}))$ and amplitude ratio $({{\hat \eta}_g^{+}/{\hat \chi}_c^{-}})$ are plotted in figure \ref{fig:6a}. 
We can explicitly see how the region of instability is bounded by the ability of the waves to be phase locked to form a growing normal mode with mutual wave amplification. 
The mean frequency ratio ${{\overline \Omega}_{R}} /{{\overline \Omega}_{L}} = (L/R)^2 <1$. Hence, for small $(L/R)$ ratio the difference between the mean flow phase speeds are relatively large thus the wave should help each other to counter-propagate against the shear to remain phase-locked $(\pi > \Delta \epsilon > \pi/2)$. As the ratio $(L/R)$ becomes closer to unity the difference between ${\Omega}_{R}$ and ${\Omega}_{L}$ decreases and the waves should hinder each others phase propagation to remain phase-locked $(\pi/2 > \Delta \epsilon > 0)$, where the margins of instability are obtained when  $\Delta \epsilon = (0,\pi)$. The most unstable mode is manifested when the phase difference is almost at  $\Delta \epsilon =0.48\pi \approx \pi/2$, which is very near to the optimal configuration for instantaneous growth when $\omega_g^{+} = \omega_c^{-} $. 

The amplitude ratio for the most unstable mode (${{\hat \eta}_g^{+}/{\hat \chi}_c^{-}}= 0.67$). However, since both $L/D<1$ and $L/R<1$, the visual effect of this amplitude ratio may  diminish the actual amplitude ratio even more. 


\subsection{The near resonance regime}

Equation set \eqref{eq36}, together with the understanding of the wave propagation and interaction mechanisms, indicate that instantaneous amplification is possible either between a pair of two counter-propagating waves $(\chi_c^{-}, { \eta}_g^{+})$, or between a pair of two pro-propagating waves  $(\chi_c^{+}, { \eta}_g^{-})$. Obviously, as was shown in the previous subsection, out of the two pairs only the growth between the counter-propagating ones can be sustained to form modal structure with exponential growth. Nevertheless one can obtain neutral normal modes which are composed off a combination of a counter and a pro propagating waves. This is the case described in figure \ref{fig:7} which includes the information of figure 2d in TMBF. The interaction is between the waves of $(\chi_c^{-}, { \eta}_g^{-})$, so that relatively to the shear the centrifugal wave $\chi_c^{-}$ is counter-propagating whereas ${\hat \eta}_g^{-}$ is pro-propagating. The waves are phase-locked to propagate in concert either when they help each other propagation when their amplitudes are in phase ($\Delta \epsilon = \epsilon_g^{-} - \epsilon_c^{-}=0)$ (figures \ref{fig:8}a,b), or when their amplitudes are in anti-phased ($\Delta \epsilon = \pi)$ (figures \ref{fig:8}c,d) when they hinder each other. Since the self propagation rate of both waves is clockwise (i.e. negative) ``helping'' will decrease the phased locked frequency and ``hindering'' will increase it. The difference in the helping and  hindering scenarios is in the amplitude ratio between the two waves. At any rate since the waves are neutral their structure will not emerge in the rotating tank experience. For completeness, however, we present here the analysis of this neutral ``near resonance'' interaction.

Writing:
\be 
\label{eq43}
\chi_c^{-} = {\hat \chi}_c^{-}(t)e^{i\[m\theta + \epsilon_c^{-}(t)\]}\,  ; \hspace{0.25cm}
\eta_g^{-} = {\hat \eta}_g^{-}(t)e^{i\[m\theta + \epsilon_g^{-}(t)\]}\, ,
\ee
and approximate \eqref{eq34} to:
\be
\label{eq44}
\(\der{}{t} + i\omega_c^{-}\)\chi_c^{-} \approx i\sigma_T\eta_g^- \,  ; \hspace{0.25cm}
\(\der{}{t} + i\omega_g^{-}\)\eta_g^{-} \approx i\sigma_B\chi_c^-  \, ,
\ee
the real parts of \eqref{eq44} can be written as:
\be
\label{eq45}
{{\dot  {\hat \chi}}_c^{-}\over {\hat \chi}_c^{-}} = -{{\hat \eta}_g^{-}\over{\hat \chi}_c^{-}}\sigma_T\sin{\Delta \epsilon}\,  ; \hspace{0.25cm}
{{\dot  {\hat \eta}}_g^{-}\over  {\hat \eta}_g^{-}} = {{\hat \chi}_c^{-}\over {\hat \eta}_g^{-}}\sigma_B\sin{\Delta \epsilon}\, .
\ee 
Hence the only modal solution is of zero growth rate:
\be
\label{eq46}
{\omega_{NM}}_i = {{\dot  {\hat \chi}}_c^{-}\over {\hat \chi}_c^{-}} = {{\dot  {\hat \eta}}_g^{-}\over  {\hat \eta}_g^{-}} = 0
\ee
yielding $\Delta \epsilon = (0,\pi)$. The imaginary parts of \eqref{eq44} become:
\be
\label{eq47}
-{\dot \epsilon}_c^{-} = \omega_c^{-} - {{\hat \eta}_g^{-}\over{\hat \chi}_c^{-}}\sigma_T\cos{\Delta \epsilon}\,  ; \hspace{0.25cm}
-{\dot \epsilon}_g^{-} = \omega_g^{-} - {{\hat \chi}_c^{-}\over {\hat \eta}_g^{-}}\sigma_B\cos{\Delta \epsilon}\, ,
\ee
thus for modal solution, ${\omega_{NM}}_r = -{\dot \epsilon}_c^{-} = -{\dot \epsilon}_g^{+}$,
\be
\label{eq48}
\Delta \omega = \pm\(\sigma_T\lambda - {\sigma_B\over \lambda}\),
\ee
with the frequency difference $\Delta \omega \equiv \(\omega_c^{-} - \omega_g^{-}\)$ and amplitude ratio $\lambda \equiv {{\hat \eta}_g^{-}/{\hat \chi}_c^{-}}$. The plus solution corresponds to $\Delta \epsilon = 0$, i.e. the ``helping'' solution, whereas the minus corresponds to the $\Delta \epsilon = \pi$, ``hindering'' one.
This gives the amplitude ratio solution:
\be
\label{eq49}
\lambda_{(help,hinder)} = {{\pm}\Delta \omega + \sqrt{(\Delta \omega)^2 +4\sigma_T\sigma_B} \over 2\sigma_T}, 
\ee
corresponding to: 
\be
\label{eq50}
{{\omega_{NM}}_r}_{(help,hinder)} = {1\over 2}\(\omega_c^{-} + \omega_g^{-}\) \mp {\sqrt{\sigma_T\sigma_B}\over 2}
\[\(\sqrt{\sigma_T \over \sigma_B}\lambda\) + \(\sqrt{\sigma_T \over \sigma_B}\lambda\)^{-1} \].
\ee
These two modal frequencies are plotted in figure \ref{fig:7} together with $\omega_c^{-}$, $\omega_g^{-}$,  $\(\omega_c^{-} + \omega_g^{-}\)/2$ and the two solutions of $\(\sqrt{\sigma_T \over \sigma_B}\lambda\)$.
$\Delta \omega  > 0$ at $L/R < 0.778$ and $\Delta \omega  < 0$ at $L/R > 0.778$. For $\Delta \omega  > 0$ phase locking can be achieved when the waves help each other to propagate
if ${\hat \eta}_g^{-}> {\hat \chi}_c^{-} $ (figure \ref{fig:8b}). Then the $g^{-}$ wave affects the $c^{-}$ one more than vice-verse thus $g^{-}$ helps $c^{-}$ to propagate more than $c^{-}$ helps 
$g^{-}$. The result is that $g^{-}$ enforces $c^{-}$ to propagate with a frequency which is even smaller than $\omega_g^{-}$ since the $c^{-}$ decreases by little the wave propagation frequency of $g^{-}$ (recall that the self propagation rate of each wave in isolation is clockwise, i.e. negative). Consequently the phase-locked normal mode frequency ${\omega_{NM}}_{help} < \omega_g^{-}$ (but still positive due to the advection of the mean flow). The other option for phase locking for $\Delta \omega  > 0$ is that the waves hinder each other and  ${\hat \chi}_c^{-}> {\hat \eta}_g^{-}$(figure \ref{fig:8c}). Then $c^{-}$ slows (thus makes the frequency more positive) $g^{-}$ more than vice-verse and enforces $g^{-}$ to propagate with a frequency which is even larger than $\omega_c^{-}$ as the $g^{-}$ increases by little the wave propagation speed of $c^{-}$. Consequently the phase-locked normal mode frequency 
${\omega_{NM}}_{hinder} > \omega_c^{-}$ (taking the advection of the mean flow as well into the account). When $\Delta \omega  < 0$ at $L/R > 0.778$ the roles between the two waves are switched. The helping and hindering configurations for this case are plotted in figures \ref{fig:8}(a,d), respectively.  
 
\begin{figure}%
    \centering
    \subfloat[]{{\fbox{\includegraphics[width=0.9\textwidth]{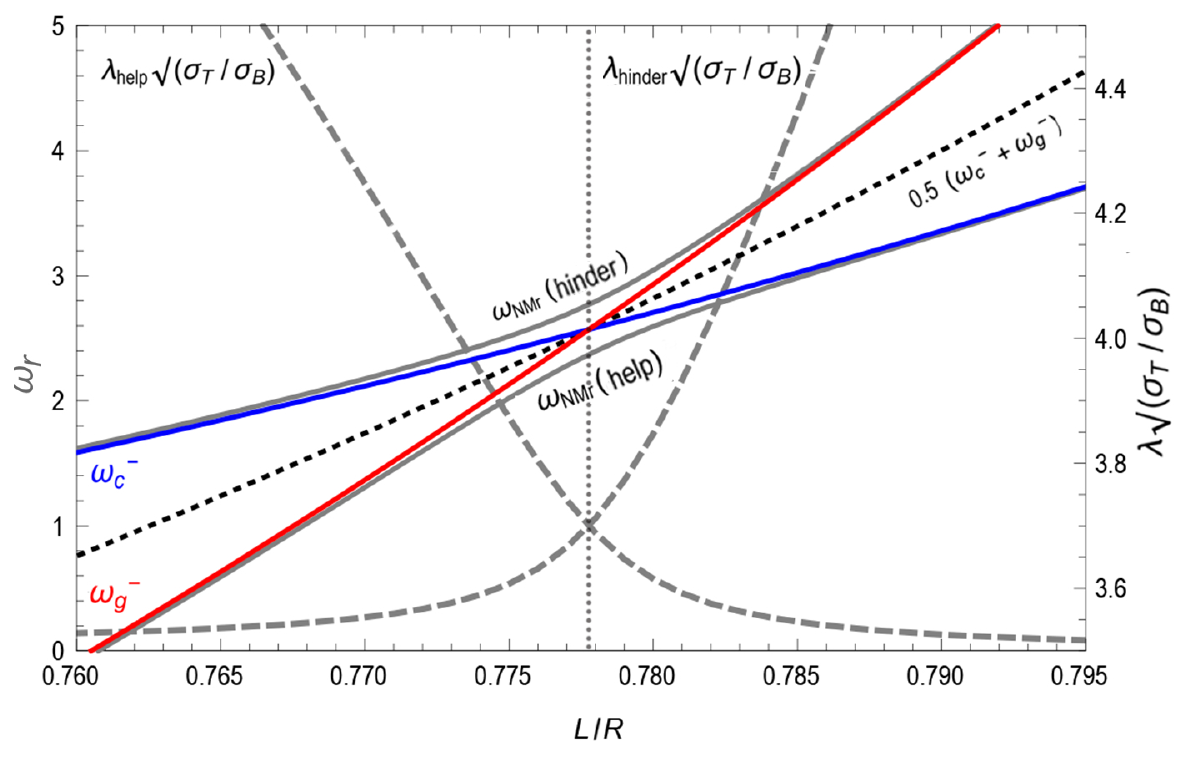}}
    \label{fig:7} }}
    \caption{\small Interaction of counter and  pro propagating waves which result in neutral normal modes, for ${H}/{R}=0.276$ and m=3. (a) The two modal frequencies (gray lines) for which the waves hinder or help the other to propagate. The wave frequencies in isolation ($\omega_c^{-}$ (blue), $\omega_g^{-}$ (red)) and their averaged  values $\(\omega_c^{-} + \omega_g^{-}\)/2$ (doted line) are shown as well for comparison. Plotted in dashed  are the scaled amplitude ratio $\(\sqrt{\sigma_T \over \sigma_B}\lambda\)$ for the helping and hindering solutions of Eq.\eqref{eq49}(right axis for coresponding values).}%
    \label{fig:7}%
\end{figure}

\begin{figure}%
    \centering
     \subfloat[]{{\fbox{\includegraphics[width=0.55\textwidth]{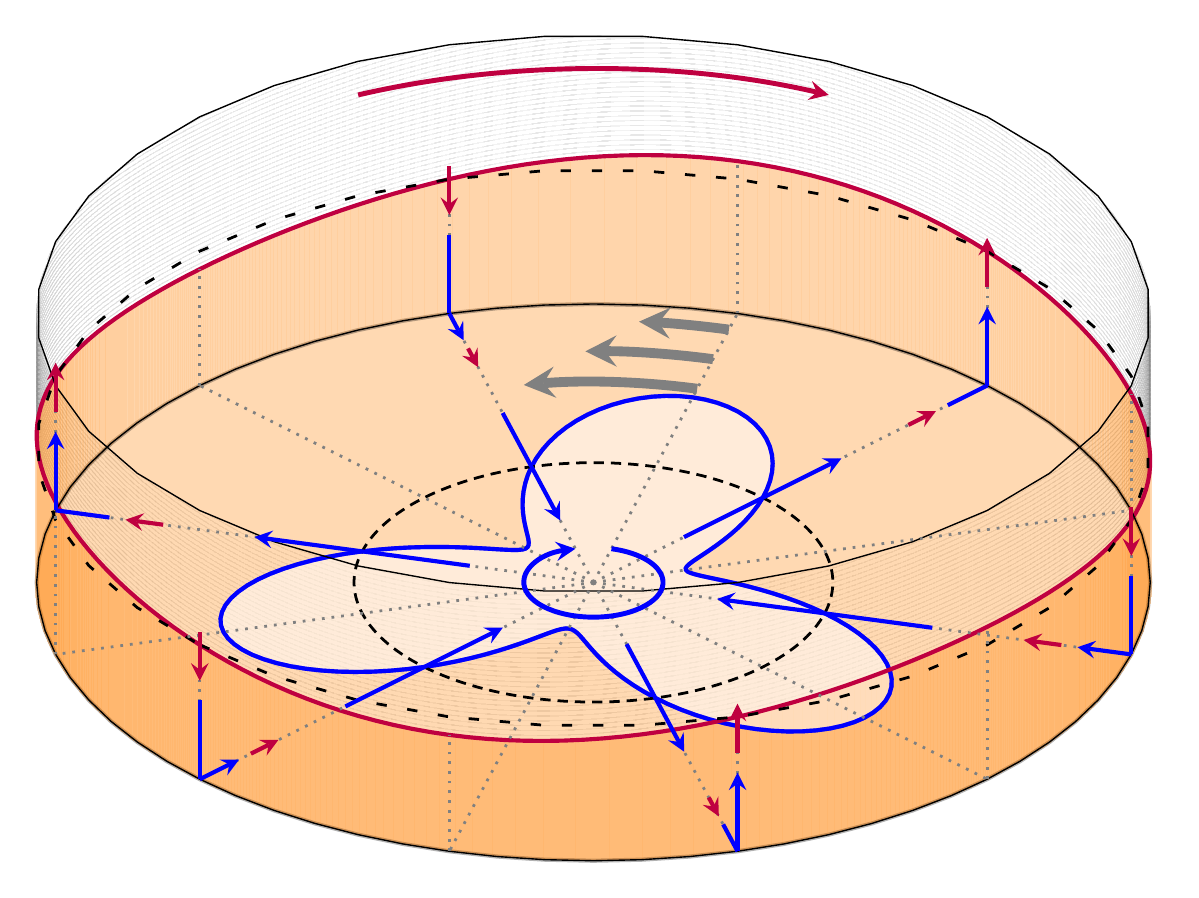}}
    \label{fig:8a} }}
    \qquad
    \subfloat[]{{\fbox{\includegraphics[width=0.55\textwidth]{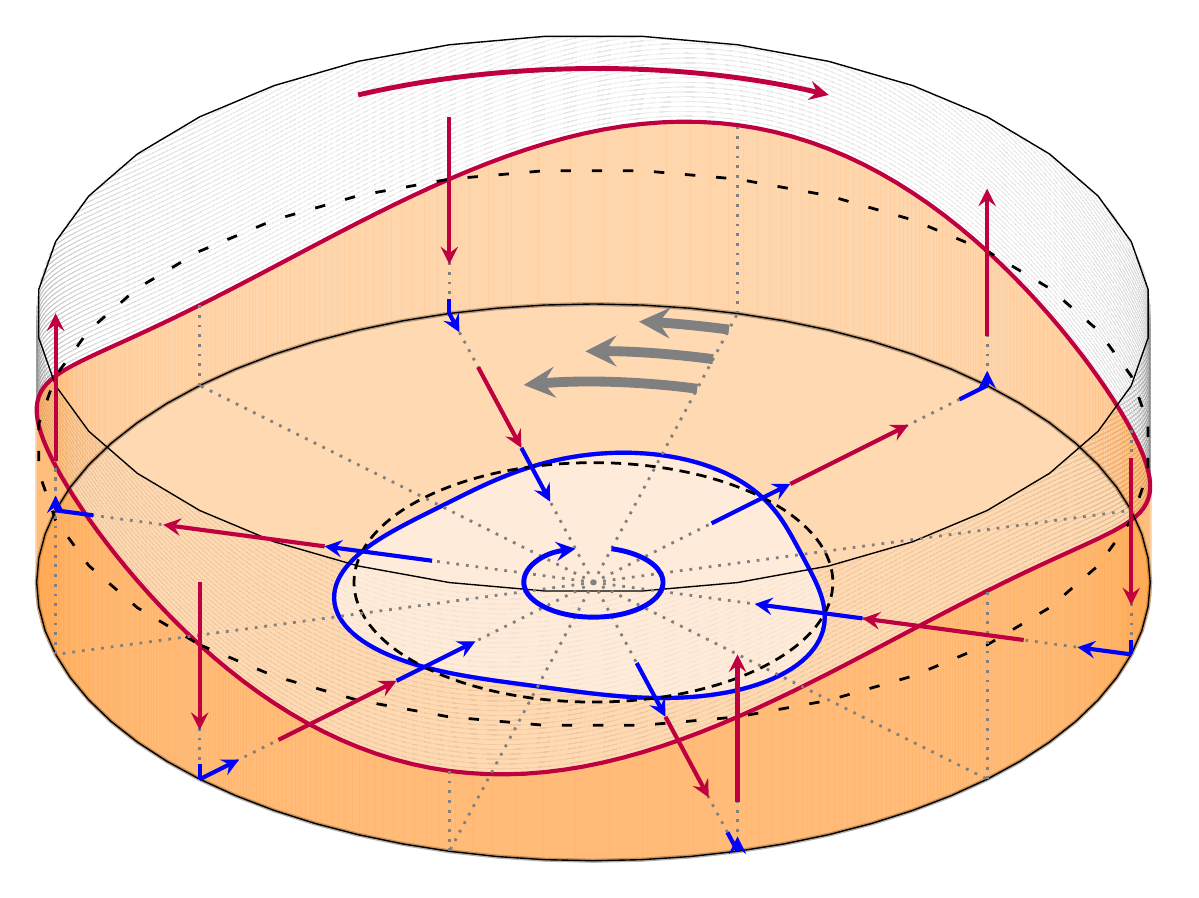}}
    \label{fig:8b} }}
    \qquad
   \subfloat[]{{\fbox{\includegraphics[width=0.55\textwidth]{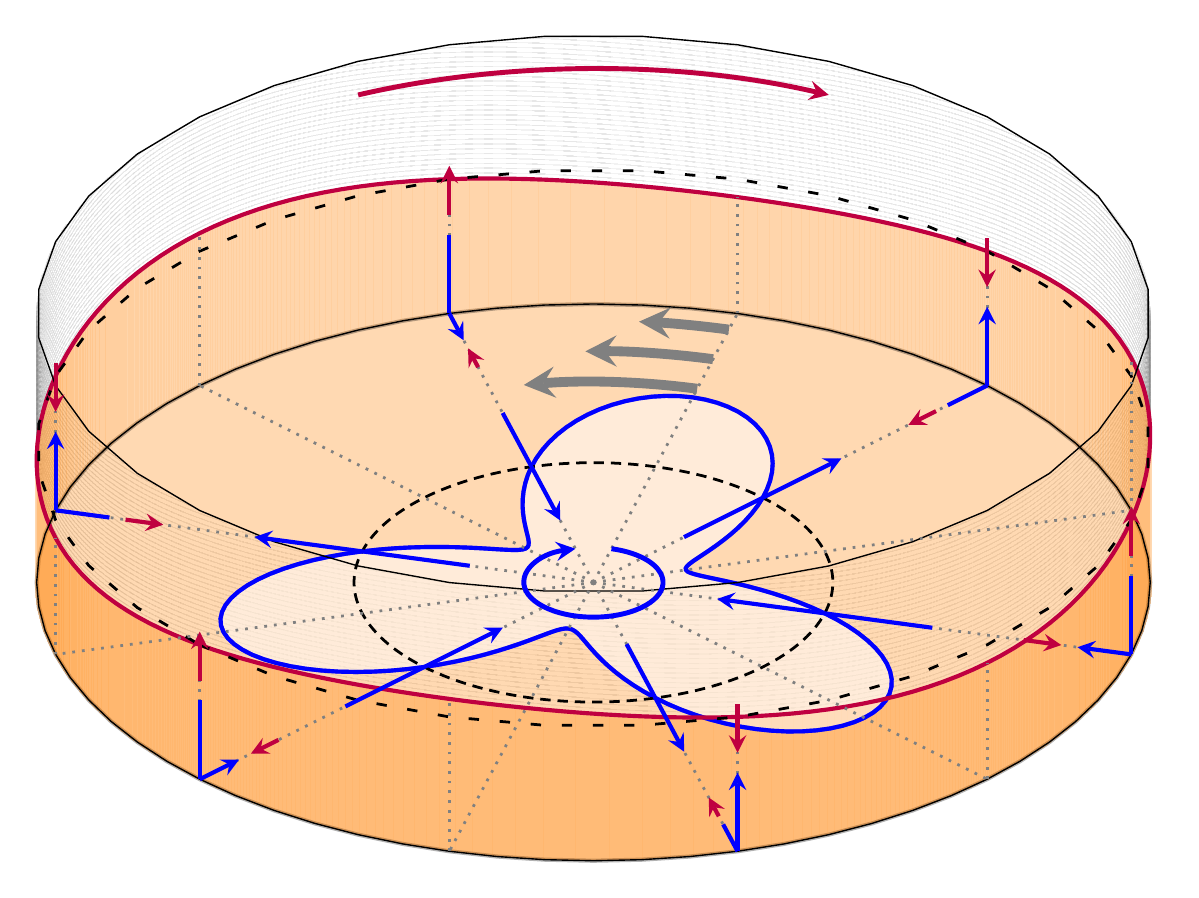}}
    \label{fig:8c} }}
    \qquad
\subfloat[]{{\fbox{\includegraphics[width=0.55\textwidth]{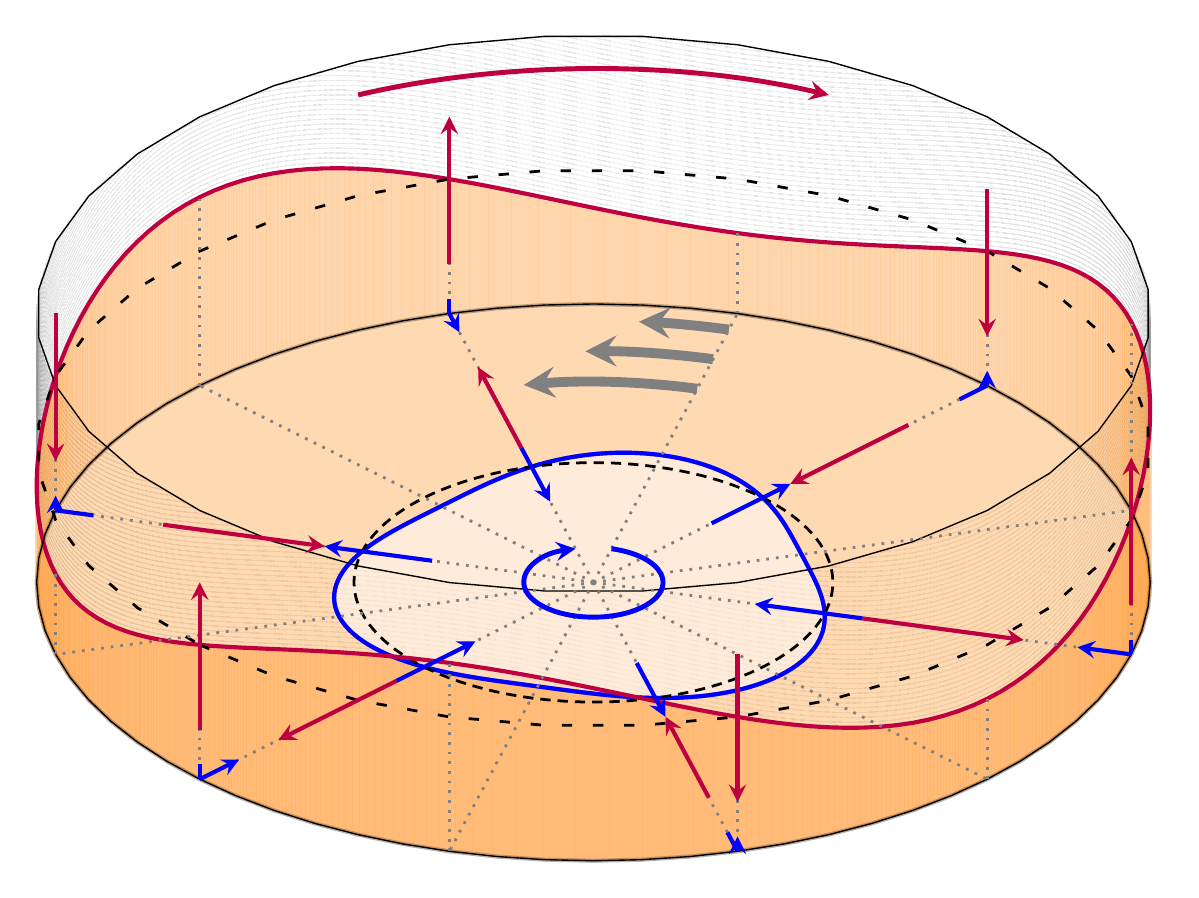}}
    \label{fig:8d} }}
    \qquad
    \caption{\small The ``near resonance'' interaction between the waves of $(\chi_c^{-}, { \eta}_g^{-})$.  The waves help each other to propagate when 
		(a) ${\hat \chi}_c^{-} > {\hat \eta}_g^{-}$ for  $L/R < 0.778$, or when  (b)${\hat \chi}_c^{-} < {\hat \eta}_g^{-}$ for$L/R > 0.778$. The waves hinder each other to propagate when   
		(c) ${\hat \chi}_c^{-} > {\hat \eta}_g^{-}$ for $L/R > 0.778$  or when (d) ${\hat \chi}_c^{-} < {\hat \eta}_g^{-}$ for $L/R <0.778$}%
    \label{fig:8}%
\end{figure}

\section{Summary}

The motivation for this study was twofold: to understand the nature of the inner surface centrifugal waves and how they form resonant instability with the outer vertical gravity waves. 
We found that the centrifugal waves are a potential flow representation of baroclinic edge waves, where the sharp gradients of density and pressure across the free fluid-air interface yield baroclinic torques which generate a wavy vortex sheet at the edge (which, in principle, is the same wave propagation mechanism discussed in \citet{heifetz2015stratified}). Depicting the dynamics as arising from the action-at-a-distance interaction between
gravity waves and centrifugal waves represented as interfacial edge wave vortex sheets (e.g. \citet{harnik2008buoyancy}) 
has allowed us to rationalize the resonant instability patterns observed in swirling flows inside fast rotating cylindrical containers.

In geophysical fluid dynamics this is done by implementing vorticity inversion of Rossby edge waves to obtain the velocity field that each wave induces on the opposed one.
The instability is then explained as a resonance between two counter-propagating Rossby waves \citep{2004QJRMS.130..211H} which are phased-locked in a growing configuration when ``the induced velocity field of each Rossby wave keeps the other in step, and makes the other grow.'' \citep{hoskins1985use} .
Here the mechanism is essentially the same however the setup is more complex. On each interface
there exist two waves rather than one and the mechanism of vorticity propagation is less straightforward than advection of mean vorticity. Furthermore, the centrifugal and the gravity waves are located on perpendicular surfaces. Nevertheless, it is shown that instability is obtained by phase locking resonance between the two counter-propagating vorticity waves (one is centrifugal and the other is gravity) and the induced velocity fields act both to phase lock the waves to propagate with the same frequency and to amplify each other amplitudes. 
This approach also explains why only counter-propagating vorticity waves can form resonant instability. Modal neutral phase-locking can be obtained in certain conditions between one pro and one counter propagating waves, however such two waves cannot result in mutual amplification. The near resonance instability regime, found in TMBF, is an example of such interaction that can lead to phase locking but not to instability.   

The vorticity wave interaction approach has been applied this far to resonant instability between Rossby \citep{heifetz1999counter}, gravity  \citep{carpenter2011instability}, capillary 
\citep{2015PhFl...27d4104B}, and even Alfven waves in shear  dynamics of plasma \citep{heifetz2015interacting}. Hence, it is our aim to analyze the other resonant instability mechanisms obtained in swirling flow experiments with lower rotation rates, where both Rossby and inertial waves participate in the resonant instability mechanism.\\

{\noindent \bf Acknowledgments}\\
EH is grateful to Tomas Bohr for showing him the experience of fast rotating swirling flow in his lab in DTU.   
 
 \bibliographystyle{apalike}

\bibliography{scibib}

\newcommand{\noop}[1]{} \newcommand{\bibstar}{{\Large $\ast$}}
  \newcommand{\noopt}[1]{} \newcommand{\bibbu}{{\Large $\bullet$}}
\begin{thebibliography}{}

\bibitem[Baines and Mitsudera, 1994]{baines1994mechanism}
Baines, P.~G. and Mitsudera, H. (1994).
\newblock On the mechanism of shear flow instabilities.
\newblock {\em Journal of Fluid Mechanics}, 276:327--342.

\bibitem[{Biancofiore} et~al., 2015]{2015PhFl...27d4104B}
{Biancofiore}, L., {Gallaire}, F., and {Heifetz}, E. (2015).
\newblock {Interaction between counterpropagating Rossby waves and capillary
  waves in planar shear flows}.
\newblock {\em Physics of Fluids}, 27(4):044104.

\bibitem[Carpenter et~al., 2011]{carpenter2011instability}
Carpenter, J.~R., Tedford, E.~W., Heifetz, E., and Lawrence, G.~A. (2011).
\newblock Instability in stratified shear flow: Review of a physical
  interpretation based on interacting waves.
\newblock {\em Applied Mechanics Reviews}, 64(6):060801.

\bibitem[Fabre and Mougel, 2014]{fabre2014generation}
Fabre, D. and Mougel, J. (2014).
\newblock Generation of three-dimensional patterns through wave interaction in
  a model of free surface swirling flow.
\newblock {\em Fluid Dynamics Research}, 46(6):061415.

\bibitem[Harnik et~al., 2008]{harnik2008buoyancy}
Harnik, i., Heifetz, E., Umurhan, O., and Lott, F. (2008).
\newblock A buoyancy-vorticity wave interaction approach to stratified shear
  flow.
\newblock {\em Journal of the Atmospheric Sciences}, 65(8):2615--2630.

\bibitem[Heifetz and Mak, 2015]{heifetz2015stratified}
Heifetz, E. and Mak, J. (2015).
\newblock Stratified shear flow instabilities in the non-boussinesq regime.
\newblock {\em Physics of Fluids (1994-present)}, 27(8):086601.

\bibitem[Heifetz et~al., 2015]{heifetz2015interacting}
Heifetz, E., Mak, J., Nycander, J., and Umurhan, O. (2015).
\newblock Interacting vorticity waves as an instability mechanism for
  magnetohydrodynamic shear instabilities.
\newblock {\em Journal of Fluid Mechanics}, 767:199--225.

\bibitem[Heifetz et~al., 1999]{heifetz1999counter}
Heifetz, i.~E., Bishop, C., and Alpert, P. (1999).
\newblock Counter-propagating rossby waves in the barotropic rayleigh model of
  shear instability.
\newblock {\em Quarterly Journal of the Royal Meteorological Society},
  125(560):2835--2853.

\bibitem[{Heifetz} et~al., 2004]{2004QJRMS.130..211H}
{Heifetz}, i.~E., {Bishop}, C.~H., {Hoskins}, B.~J., and {Methven}, J. (2004).
\newblock {The counter-propagating Rossby-wave perspective on baroclinic
  instability. I: Mathematical basis}.
\newblock {\em Quarterly Journal of the Royal Meteorological Society},
  130:211--231.

\bibitem[Hoskins et~al., 1985]{hoskins1985use}
Hoskins, i.~J., McIntyre, M., and Robertson, A.~W. (1985).
\newblock On the use and significance of isentropic potential vorticity maps.
\newblock {\em Quarterly Journal of the Royal Meteorological Society},
  111(470):877--946.

\bibitem[Mougel et~al., 2014]{mougel2014waves}
Mougel, J., Fabre, D., and Lacaze, L. (2014).
\newblock Waves and instabilities in rotating free surface flows.
\newblock {\em Mechanics \& Industry}, 15(2):107--112.

\bibitem[Mougel et~al., 2015]{mougel2015waves}
Mougel, J., Fabre, D., and Lacaze, L. (2015).
\newblock Waves in newton?s bucket.
\newblock {\em Journal of Fluid Mechanics}, 783:211--250.

\bibitem[Toph{\o}j et~al., 2013]{tophoj2013rotating}
Toph{\o}j, L., Mougel, J., Bohr, T., and Fabre, D. (2013).
\newblock Rotating polygon instability of a swirling free surface flow.
\newblock {\em Physical review letters}, 110(19):194502.

\end{thebibliography}
\end{document}